\documentclass[12pt,preprint]{aastex}

\slugcomment{Accepted for publication in AJ tentatively in the March issue}

\begin{document}

\newcommand {\hi} {\ion{H}{1}\,\,}
\newcommand {\hii} {\ion{H}{2}\,\,}
\newcommand {\msol} {M$_\odot$\,}
\newcommand {\lsol} {L$_\odot$\,}
\newcommand {\mhil} {$M_{HI} /  L^{0}_B$\,}
\newcommand {\mlsol} {M$_\odot$/L$_\odot$\,}

\title{The Recent Star Formation History of NGC 5102\altaffilmark{1}} 

\author{Sylvie F. Beaulieu}
\affil{D\'epartement de Physique, de G\'enie Physique et d'Optique
and Centre de Recherche en Astrophysique du Qu\'ebec (CRAQ),
Universit\'e Laval, Qu\'ebec, QC, G1V 0A6, Canada}
\email{sbeaulieu@phy.ulaval.ca}

\author{Kenneth C. Freeman}
\affil{Research School of Astronomy and Astrophysics (RSAA), Australian National University,
Cotter Road, Weston Creek, ACT 2611, Australia}
\email{kcf@mso.anu.edu.au}

\author{Sebastian L. Hidalgo}
\affil{Instituto de Astrof\'isica de Canarias, Tenerife, Spain}
\email{shidalgo@iac.es}

\author{Colin A. Norman}
\affil{Center for Astrophysical Sciences, Department of Physics and Astronomy, 
The Johns Hopkins University, Baltimore, MD 21218, USA}
\email{norman@stsci.edu}

\and

\author{Peter J. Quinn}
\affil{ICRAR, University of Western Australia, Crawley, WA, 6009, Australia}
\email{peter.quinn@uwa.edu.au}

\altaffiltext{1}{Based on observations with the NASA/ESA {\it Hubble
Space Telescope}, obtained at the Space Telescope Science Institute,
which is operated by the Association of Universities for Research in
Astronomy (AURA), Inc., under NASA Contract NAS 5-26555.}

\begin{abstract}
We present {\it Hubble Space Telescope} photometry of young stars in NGC 5102, 
a nearby gas-rich 
post-starburst S0 galaxy with a bright young stellar nucleus. We use the 
IAC-pop/MinnIAC algorithm to derive the recent star formation history in three 
fields in the bulge and disk of NGC 5102. In the disk fields, the recent star 
formation rate has declined monotonically and is now barely detectable, but 
a starburst is still in progress in the bulge and has added about 2\%
to the mass of the bulge over the last 200 Myr. Other studies of star formation 
in NGC 5102 indicate that about 20\% of its stellar mass was added over the 
past Gyr. If this is correct, then much of the stellar mass of the bulge may have 
formed over this period. It seems likely that this star formation was fueled by 
the accretion of a gas-rich system with \hi mass of about $2 \times 10^9$ \msol\
which has now been almost completely converted into stars. The large mass of 
recently formed stars and the blue colours of the bulge suggest that the current 
starburst, which is now fading,  may have made a significant contribution to 
build the bulge of NGC 5102. 
\end{abstract}


\section{Introduction}

NGC 5102 is a low-luminosity galaxy of classic S0 appearance in the nearby
Cen A group, at a distance of about 3.5 Mpc (van den Bergh 1976). Slightly 
smaller distances have been estimated by Karachentsev et al (2002) and
Davidge (2008a), who note the potential effects of intermediate-age populations 
on the tip of the red giant distance. Table \ref {N5102-param} summarises its basic 
parameters. Although it has the disk/bulge structure of a normal S0 galaxy (Pritchet 
1979), NGC 5102 contains about $3 \times 10^8$ \msol of \hi (van Woerden et al.
1993\footnote{Values from the van Woerden et al. paper have been adjusted 
to our adopted distance through this paper.}), an unusually bright nucleus 
dominated by young stars in optical light (Gallagher et al. 1975, van den Bergh 
1976), and an extended distribution of ionized gas and dust in the inner regions 
of its bulge (McMillan et al 1994, Danks et al. 1979, Xilouris et al. 2004). A
sprinkling of resolved stars from recent star formation is seen throughout its disk
(van den Bergh 1976, Davidge 2008a).

It is not yet clear whether NGC 5102 is simply an S0 galaxy that has undergone a
recent episode of star formation, or it is making a transition to the S0 state 
from a state more like a low-luminosity spiral (or vice versa; van Woerden et al.
1993; Davidge, 2008a). The formation of S0 galaxies is still poorly understood. We do 
not know how disk galaxies lose their interstellar medium to make the transition to 
the S0 state. With its S0 appearance and continuing low level of star formation, 
NGC 5102 may be a system caught in this transition phase, so we would like to 
understand the implications of its gas distribution and recent star formation history (SFH). 
One element in solving this problem is to delineate the recent SFH
more precisely. Our main goal in this paper is to derive the history and amount of 
recent star formation in different regions of NGC 5102, particularly in the inner regions 
where the effects of the recent star formation are most visible.

The blue stellar nucleus of NGC 5102, with an absolute magnitude of $M_V$ = -14.1, 
is significantly brighter than the typical nuclei of spiral galaxies (e.g., B\"oker et al. 2004). 
{\it IUE} spectra confirm the presence of early-type stars (Rocca-Volmerange \& Guiderdoni 
1987). The nucleus has an A-type absorption spectrum (Gallagher et al. 1975) and its 
colours ({\it B-V} = 0.2 and {\it U-B} = -0.1) are consistent with a single burst system of age $10^{8}$ 
years and a mass of about $6 \times 10^6$ \msol (Pritchet 1979). Our {\it Hubble Space Telescope} 
({\it HST}) optical (WFPC2) and near-IR (NICMOS) images (to be discussed elsewhere) show that the 
nucleus is spatially unresolved: its FWHM is less than 3 pc. This is a compact configuration 
for such a bright stellar nucleus. We note that there is no evidence for current active galactic nucleus 
(AGN)-type nuclear activity from near-IR, optical, UV or X-ray spectra (Bendo \& Joseph 2004; Gallagher 
et al 1975; Chandar et al 2004; Irwin et al. 2004).

Away from the nucleus, the {\it B-V} colours of the bulge and disk show a weak radial 
gradient, with {\it B-V} increasing with radius from about 0.65 in the bulge to about 0.75 
in the disk (Pritchet 1979). The photometry of Sandage \& Visvanathan (1978) shows 
that the mean $b-V$ and $V-r$ colours of the disk and bulge of NGC 5102 (between
circular apertures of 15\arcsec\ and 120\arcsec) are similar to the mean colours of 
Virgo S0 galaxies. The $u-V$ colour in this region of NGC 5102 is however at the blue 
extreme for these systems. These colours indicate the recent star formation superimposed 
on an older disk population. It appears that the effects of the recent star formation episode 
are widespread throughout NGC 5102, but are strongest in the inner regions. Direct 
evidence for an old population of stars in the outer disk comes from Davidge's (2008a) 
detection of red giants (see also Karachentsev et al 2002).

The \hi in NGC 5102 is distributed in a ring of mass $2 \times 10^8$ \msol, radius 
3.5 kpc and peak surface density $1.5$ \msol pc$^{-2}$, with a central depression; 
the \hi surface density rises again to about $0.9$ \msol pc$^{-2}$ near its centre 
(van Woerden et al. 1993). Within this central depression, McMillan et al. (1994)
detected a prominent filament of ionized gas. The gas shows emission lines of [OI], 
[OII] and O[III], and is therefore unlikely to be photoionized; McMillan et al. favour shock 
excitation. Furthermore, from its expansion velocity, this filament appears to be younger 
($\sim 10^7$ yr) than the nuclear stars, and McMillan et al. argued that they are probably 
not associated.

In summary, the \hi ring, the presence of extended ionized gas within it, the active recent 
star formation in the inner bulge and the relatively long time since the most recent star
formation event in the disk, suggest that we are seeing the effects of infall of gas rather 
than the burning out of a star-forming galaxy. This infall may have been in the form of a 
gas-rich galaxy, or of gas from the group environment. We note that most of the major 
galaxies in the Cen A group are disturbed in various ways (M83, Cen A, NGC 5253 and 
NGC 5102 itself) and several authors have suggested that gas from the group environment 
is responsible for these disturbances. This group gas has yet to be detected directly. A
deep \hi survey of the central regions of the Cen A group by de Blok et al. (2002) did not 
detect any new \hi clouds; this survey was sensitive to clouds with \hi masses greater than 
about $3 \times 10^6$ \msol. We also note that NGC 5102 is relatively isolated in the Cen 
A group (see Karachentsev et al 2002; Figure 1) and has no companions close enough to 
have recently interacted with it.

This paper presents the results obtained from observations using the {\it HST} WFPC2 detector 
(program 5400). We concentrate on the photometry of the PC fields only, because they were 
specifically positioned along the major axis of NGC 5102, and the PC chip, with its small pixel
size, gives adequate signal-to-noise ratio (S/N) in the high surface brightness regions of NGC 5102. 
A future paper will address the stellar content, structure and dynamics of the nucleus.

\section{WFPC2 Photometry}

Three separate pointings of NGC 5102 were obtained with WFPC2 in the {\it B}$-$ and {\it V}$-$ band 
(F450W and F569W) in 1994 September and 1995 March. Table \ref{N5102-obs} gives details of the 
observing log. The fields were centred on the PC chip with the first field F1 positioned at the centre 
of NGC 5102, and then offset twice along the SW major axis of the galaxy so as to provide a field 
F2 in a transition region between the bulge and the disk (at radius $66\arcsec$), and then a field
F3 in the disk (at $132\arcsec$). Figure \ref{DSS_3POS} shows an overlay of the three PC positions 
on a Digitized Sky Survey (DSS) {\it R} image.

A background field was also obtained in order to estimate directly the contribution of the Galactic 
background. Details on the position for each field are summarised in Table \ref{N5102-pos}.

We used the F450W and F569W filters because they provided (in Cycle 4) satisfactory transformations 
to the standard BV system and they gave the best S/N for observations of our blue stars against 
the redder diffuse background of NGC 5102.

Figures \ref{WFPC2-PCs-1}, \ref{WFPC2-PCs-2} and \ref{WFPC2-PCs-3} show the WFPC2 PC 
images in the {\it V}$-$ band of the three fields. The PC chip has 800 x 800 pixels with a scale of 
$\rm 0\arcsec.0455~pixel^{-1}$. The detector was operating at $\rm -88^\circ C$ in the low-gain 
setting of $7~e^{-}$ $\rm DN^{-1}$. Instrumental parameters and in-orbit characteristics of WFPC2 
can be found in Biretta et al. (1996).

The images were calibrated using the standard {\it HST} pipeline procedure and point-spread function 
fitting photometry 
was performed using HSTphot (Dolphin 2000a, Dolphin 2000b). The calibration includes corrections 
for charge transfer (in)efficiency (CTE), geometric distortion, aperture corrections , and zero-pointing. 
The total exposure time for each field in each filter is listed in column 5 of Table \ref{N5102-pos}.

NGC 5102 has a very bright steep core and is severely affected by dust obscuration near its centre. 
The centre of NGC 5102 is saturated in both filters of our WFPC2 data within a $0\arcsec.5$ radius. 
In an attempt to map the dust lanes and patches in F1 and to facilitate the identification of stars, we 
used the IRAF\footnote{IRAF is distributed by the National Optical Astronomy Observatories, which 
are operated by the AURA, Inc., under co-operative 
agreement with the National Science Foundation.} /STSDAS routines ELLIPSE and BMODEL in 
order to remove the background galaxy. Figure \ref{BMODEL-stars} shows the PC negative image of 
F1 with a smooth model subtracted. Results from the isophotal analysis, generated by the isophote 
fitting task ELLIPSE (Figures \ref{ELLIP-E} and \ref{ELLIP-PA}), show that the ellipticity of the galaxy 
increases gradually as the semimajor axis increases from the inner bulge to the surrounding disk, 
and the position angle of each fitted ellipses (excluding the inner $0\arcsec.5$ radius region) 
remains fairly constant. This confirms the results obtained by Williams \& Schwarzschild (1979) 
and Pritchet (1979) that NGC 5102 shows no obvious twist.

In Figure \ref{BMODEL-stars} we can clearly see for the first time, in the central region of NGC 5102 
(within $18\arcsec$ of the nucleus), the long-suspected dust lanes and patches, as well as the 
bright resolved stars from the starburst. McMillan et al. (1994) show an [OIII] $\lambda 5007$ 
``difference" image of NGC 5102 over an extended region of $3\arcmin \times 5\arcmin$. The 
filamentary [OIII] emission extends over our two inner fields, although it cannot be seen in our 
broadband WFPC2 images. Figure \ref{BMODEL-stars} also shows the stars of F1 overlaid to
illustrates our efforts to avoid using stars in the dusty regions and the messy and saturated central 
region within a radius of $0\arcsec.5$.

Colour-magnitude data for our three galaxy fields are shown in Figure \ref{CMD-PCs-Dered} 
(left panels), with their associated photometric errors (right panels). Only stars with magnitude 
errors $\leq 0.2$ and a sharpness of $\pm 0.5$ were retained. The colour-magnitude 
data for the background field F4 is shown in Figure \ref{CMD-PCs-bkgd}. As mentioned earlier, 
the background field serves to measure the contribution from the local Galactic background. Here, 
we clearly see that contamination from background stars is not a serious problem.

We applied a Galactic extinction correction to the three galaxy fields using values of 
$A_{B} = 0.237$ and $A_{V} = 0.182$ from Schlegel et al. (1998). The central field has a fair amount 
of dust, as can be seen in Figure \ref{BMODEL-stars}. We do not apply any extra reddening, as 
explained below, the uncertainty in reddening is absorbed in the derivation of the SFH, 
via a minimum $\chi^2$ criterion. Overlaid on the CMDs are a series of $z = 0.02$ Padova Isochrones 
(F450W:Girardi et al. 2002; F569W: L. Girardi 2001, private communication); log(age) is shown next 
to each isochrone.

HSTphot has a routine to estimate the completeness of our photometry. A total of 590,000 artificial
stars were distributed in the observed fields. Colours and magnitudes of the artificial stars were selected 
to sample the observed CMDs. The photometric list of artificial stars was derived exactly as for the original 
images.

The completeness as a function of colour for the $V$-band of our three fields is shown in 
Figure \ref{CompFunc}. The completeness shows a clear dependence on colour. Because of the 
high surface brightness and crowding in the inner field F1, the completeness limit is much brighter than 
in the outer fields F2 and F3.

\section{Star Formation History}

We now derive the SFH in the three fields from the stellar photometry. The 
SFH of NGC 5102 was obtained using the IAC-pop/MinnIAC algorithm (Aparicio \& Hidalgo 2009). 
The method is based on the comparison of a synthetic CMD with the observed CMD by assuming 
that any SFH can be given as linear combination of simple stellar populations (SSPs). An SSP is
defined by a set of stars with a small range in age and metallicity. The comparison is done by
dividing the synthetic CMD in partial CMD models, each one composed by an SSP. The observed 
CMD and the synthetic CMD are gridding by defining a set of boxes on them. The SFH best matching 
the distribution of the stars in the CMDs is found using a chi-squared merit function. See 
Aparicio \& Hidalgo (2009) for a detailed description of the method.

To build the synthetic CMD, we need to adopt a metallicity for NGC 5102. Davidge (2008a) derived the 
metallicity distribution function for the old red giants of the outer disk of NGC 5102;  he found that the 
metallicity ranges over the interval [M/H] = -0.9 to -0.1, with the distribution peaking near [M/H] = -0.6. 
As we will be estimating the SFH for relatively young stars, it seems appropriate to use the metallicity for 
the interstellar medium. The luminosity-metallicity relation of Tremonti et al. (2004) indicates an [O/H] 
value of about -0.2 (adopting a solar value of 12 + log(O/H) = 8.7). The compilation by Chandar et al. 
(2004) gives an [O/H] value of +0.3.  We will adopt a solar metallicity {\it Z} = 0.02.

The method of Aparicio \& Gallart (2004) was used to generate the synthetic CMD with a total of 
600,000 stars. Girardi et al. (2000) stellar evolution library and bolometric corrections from Origlia
\& Leitherer (2000) were used. We adopted an initial mass function by Kroupa et al. (1993) with 
$20 \%$ of binary stars.

Observational errors were simulated in the synthetic CMD by using the result of the completeness 
tests (Figure \ref{CompFunc}). Only stars with a magnitude error smaller than 0.2 and corrected for 
Galactic extinction are used to derive the SFH (i.e., data from Figure \ref{CMD-PCs-Dered}).

An age binning of 20 Myr was selected for stars younger than 80 Myr. For older stars, a larger 
binning was selected. The CMDs were sampled with boxes of sizes 0.04 in colour and 0.2 in 
magnitude. We used 15 sets of SSPs to minimise the effect of the age sampling in the SFH. 
The sampling boxes in the CMDs were also shifted. This produced a set of solutions which were 
averaged and the average is adopted as the final SFH. The $1\sigma$ dispersion of the averaged 
solutions are used as error bars.

To minimise the uncertainties in the stellar evolution models, photometric zero-points, reddening and 
distance estimation, the SFHs were obtained by introducing different offsets in colour and magnitude
in the CMD. The solution giving the minimum chi-squared was adopted as the best fit. This minimise
the impact in the derivation of the SFH of poorly known variables whose effect is to shift the CMD.
This is the case of the internal reddening mentioned in section 2. The technique finds the offsets
that should be introduced to the CMD to obtain the SFH with the minimum chi-squared. 

A note about the stability and uniqueness of the solution. The SFHs have been averaged by using different 
sampling parameters, which minimise the fluctuations in the SFH. This gives a solution which is stable 
independently of the age binning used. The uniqueness of the solution is related with the assumption 
made for the auxiliary functions used to build the synthetic CMD: initial mass function, metallicity, and binary 
fraction. If 
we assume that these functions are a good approximation of those present in the galaxy, then our solution 
can be consider as unique. A detailed discussion about the stability and uniqueness of the solutions 
provided by IAC-pop/MinnIAC can be found in Aparicio \& Hidalgo (2009).

Figures 11-16 present the SFH and CMDs (observed and solved) for our three fields.
The CMD of the bulge field F1 (Figure \ref {CMD-PCs-Dered}) shows a history of recent star formation. 
The isochrones in Figure \ref {CMD-PCs-Dered} show that stars as young as $10^7$ years are present.
By comparing the dereddened CMD of F1 with the synthetic CMD (Figure \ref {OBSvsSYN-F1}), we 
can make a quantitative estimate of the SFH. In Figure \ref {SFH-F1}, we show the SFH
of the central field F1. Because of the relatively bright background (and thus of the 
completeness limit), we cannot estimate the SFH in F1 for stellar ages greater than
about $200$ Myr. The star formation rate appears to have been relatively high over the past 20 Myr, in 
contrast to the two disk fields F2 and F3 (note that the vertical scales in Figures \ref {SFH-F1}, 
\ref {SFH-F2}, and \ref {SFH-F3} are not all the same).

For the inner disk field F2, we present the observed and synthetic CMDs in Figure \ref{OBSvsSYN-F2}. 
In this field, the SFH (Figure \ref {SFH-F2}) shows a decline in the star formation rate 
from about 0.5 Gyr; the past 200 Myr appear to have been quiescent. 

We show the observed and synthetic CMDs for the disk field F3 in Figure \ref {OBSvsSYN-F3}, and 
the SFH in Figure \ref {SFH-F3}. The star formation rate has declined smoothly from 
about 200 Myr ago, and has been quiescent over the past 40 Myr.

\section{Discussion and Conclusion}

We have measured the recent SFH of three fields in NGC 5102: the inner bulge, 
inner disk and disk. In the two outer fields, the SFH is qualitatively similar, showing a declining star
formation rate with no detectable star formation in the past 40 Myr. In the innermost field, the star 
formation rate has been more or less constant over the past 200 Myr, but with an abrupt burst
in the past 20 Myr. This is all consistent with the earlier data from colours and integrated spectra. 
What can we conclude about the nature of NGC 5102: is it a system that has recently made the
transition to the S0 state, or are we simply seeing the after-effects of a star formation event fueled 
by the ring of \hi that lies only a few kpc from the centre ?  How much stellar mass was involved in
this star-forming episode, and is it a significant fraction of the total stellar mass of NGC 5102 ?

First we estimate the present stellar mass of NGC 5102. This galaxy has a strong colour gradient 
(e.g. Pritchet 1979), so its {\it M/L} ratio changes significantly with radius. We adopt the B-band surface
brightness profile given by van Woerden et al (1993). When integrated over radius, after adjusting the 
surface brightness of the disk for inclination (line of sight effect only, no internal extinction correction), 
this profile gives an apparent {\it B} magnitude of 10.34, in excellent agreement with the RC3 value in 
Table \ref {N5102-param}. The disk and the bulge provide about 60\% and 40\% of the total light respectively. 
To calculate the radial change of {\it M/L}, we use the {\it B-R} data of Pritchet (1979) and Persson et al. (1979) 
which extend to a radius of about 60 arcsec, and assume that the unreddened colour ({\it B-R})$_\circ$ 
remains constant at 1.50 at larger radii. We then use the models of Bell \& de Jong (2001; excluding 
the closed box and dynamical time models which they regard as implausible) to derive the local 
{\it M/L}$_{\rm B}$ value from the ({\it B-R})$_\circ$ colour at each radius. Integrating the stellar surface density 
over the galaxy for each of the models gives a mean total stellar mass of $(5.6 \pm 0.8) \times 10^9$ \msol, 
where the error shows the scatter between the models. This is in good agreement with Davidge"s (2008a) 
dynamical estimate of $7 \times 10^9$ \msol.

We now assume that the SFH is axisymmetric, and integrate the star formation rates shown
in Figures \ref{SFH-F1}, \ref{SFH-F2}, and \ref{SFH-F3} over the region of NGC 5102, 
sampled by our three fields. For F2 and F3, we are therefore deriving the star formation rates 
integrated over annuli around NGC 5102 with the width and mean radius of these two fields. To 
integrate the mass of stars formed, we first need the mass of recently formed stars per unit area 
of the stellar disk. The observed star formation rates shown in Figures \ref{SFH-F1}, \ref{SFH-F2}, and 
\ref{SFH-F3} are per unit area on the sky, so the surface density of stars formed must be corrected 
for the inclination of NGC 5102. We derive an inclination of $66^\circ$ (face-on is $0^\circ$) using 
axial ratios measured at several isophote levels in the outer disk from the DSS R-band image of 
NGC 5102 and assuming an intrinsic axial ratio of 0.2 for the disk. The mean surface densities of 
newly formed stars in each field are given in Table \ref{N5102-sfr}: Column 1 is the field, Column 2 is 
the surface density of new stars, and Column 3 shows the total mass of stars formed over the past 
200 Myr in F1 itself and in the annuli associated with F2 and F3 as described above. Column 4 
shows the time interval over which the star formation rate was integrated.

\begin{figure}
\plotone{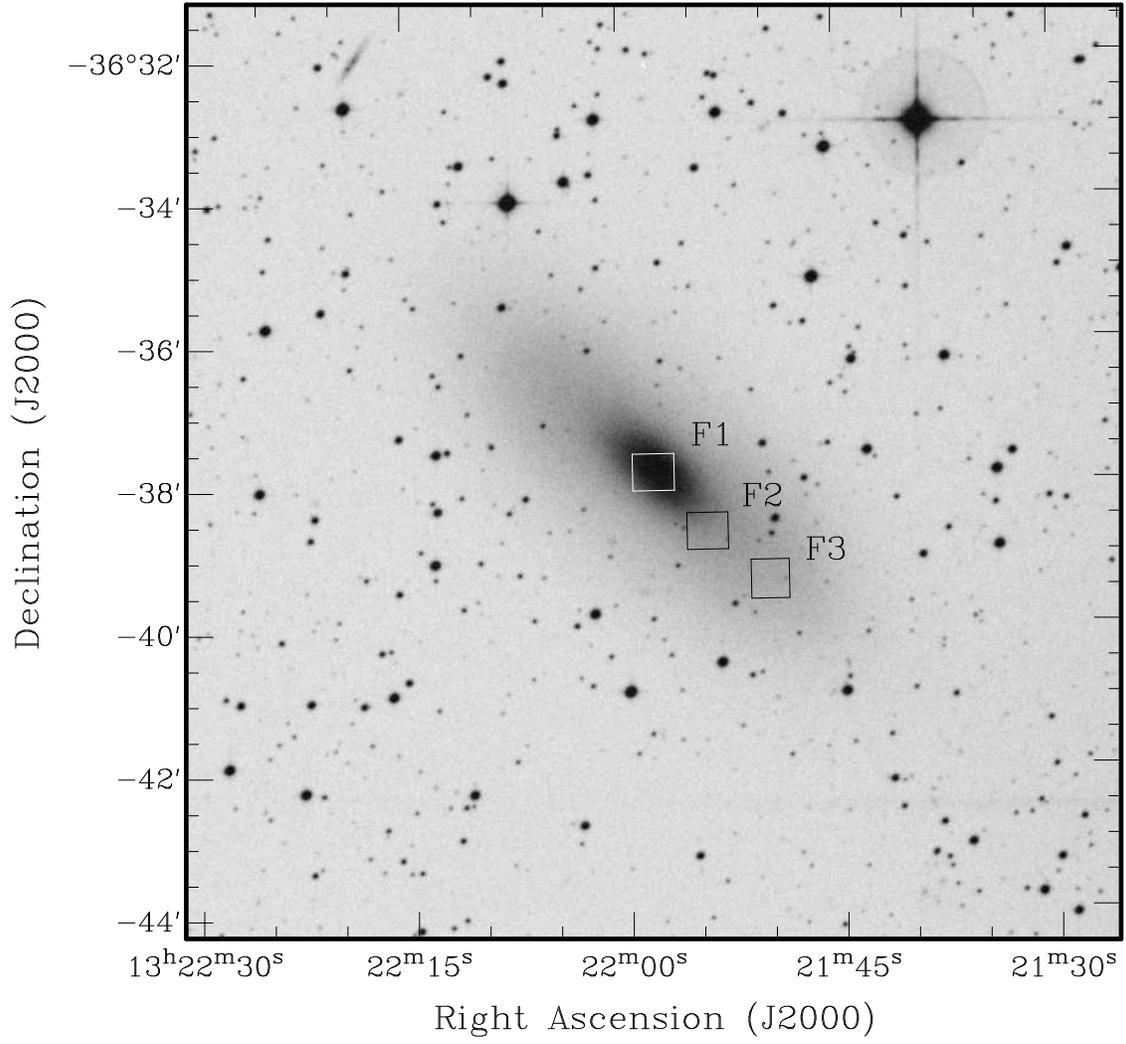}
\caption{DSS2 R-band image of NGC 5102 showing the position of the 
three WFPC2 PC fields. 
\label{DSS_3POS}}
\end{figure}

\begin{figure}
\plotone{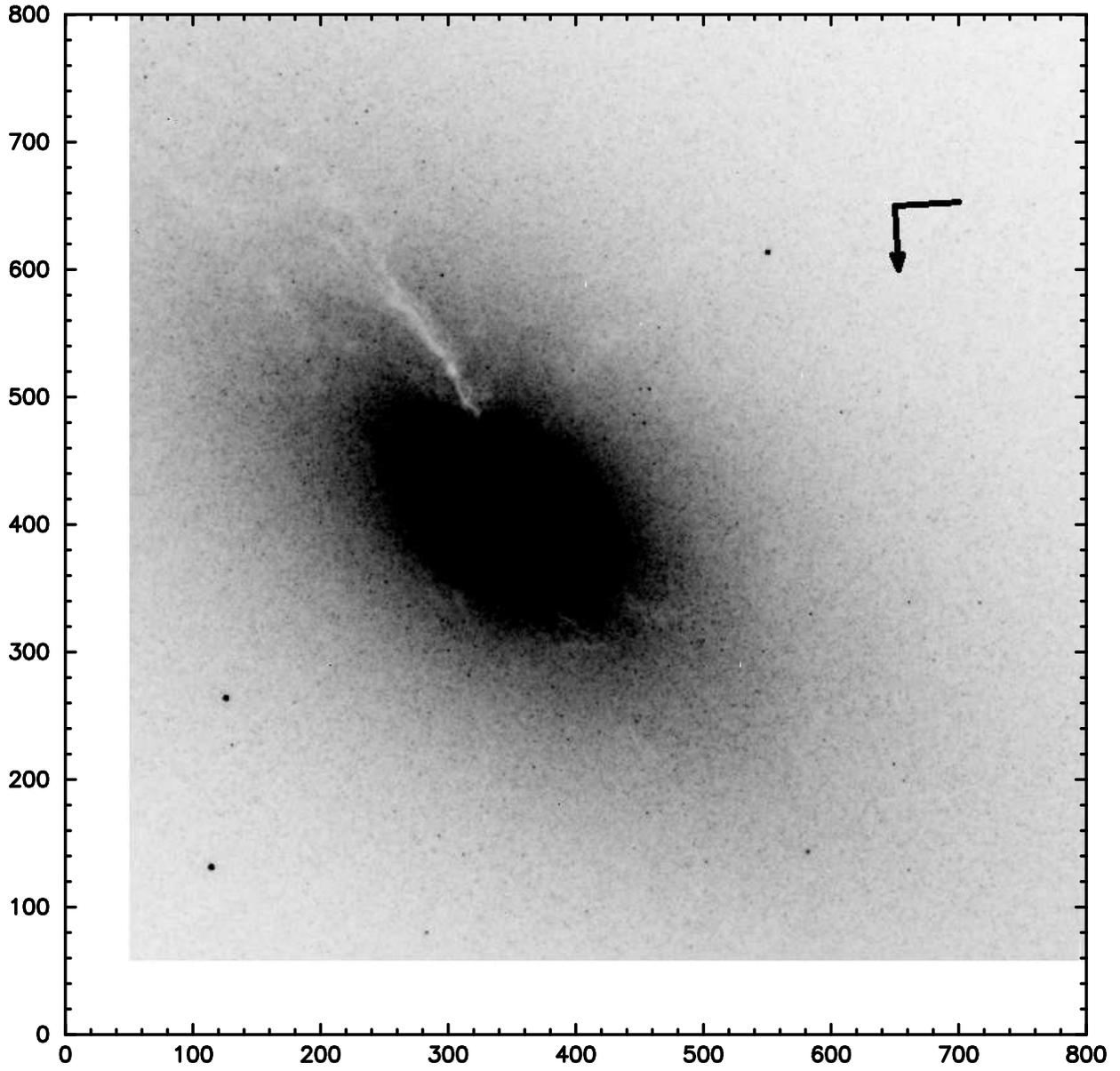}
\caption{{\it HST}/WFPC2 PC image in the $V$- band for the centre of NGC 5102 
(field 1). The arrow points to the north. The PC chip has 800 x 800 pixels with 
a scale of $\rm 0^{\arcsec}.0455~ pixel^{-1}$. The CCD characteristics are
the same for Figures 2, 3 and 4. 
\label{WFPC2-PCs-1}}
\end{figure}

\begin{figure}
\plotone{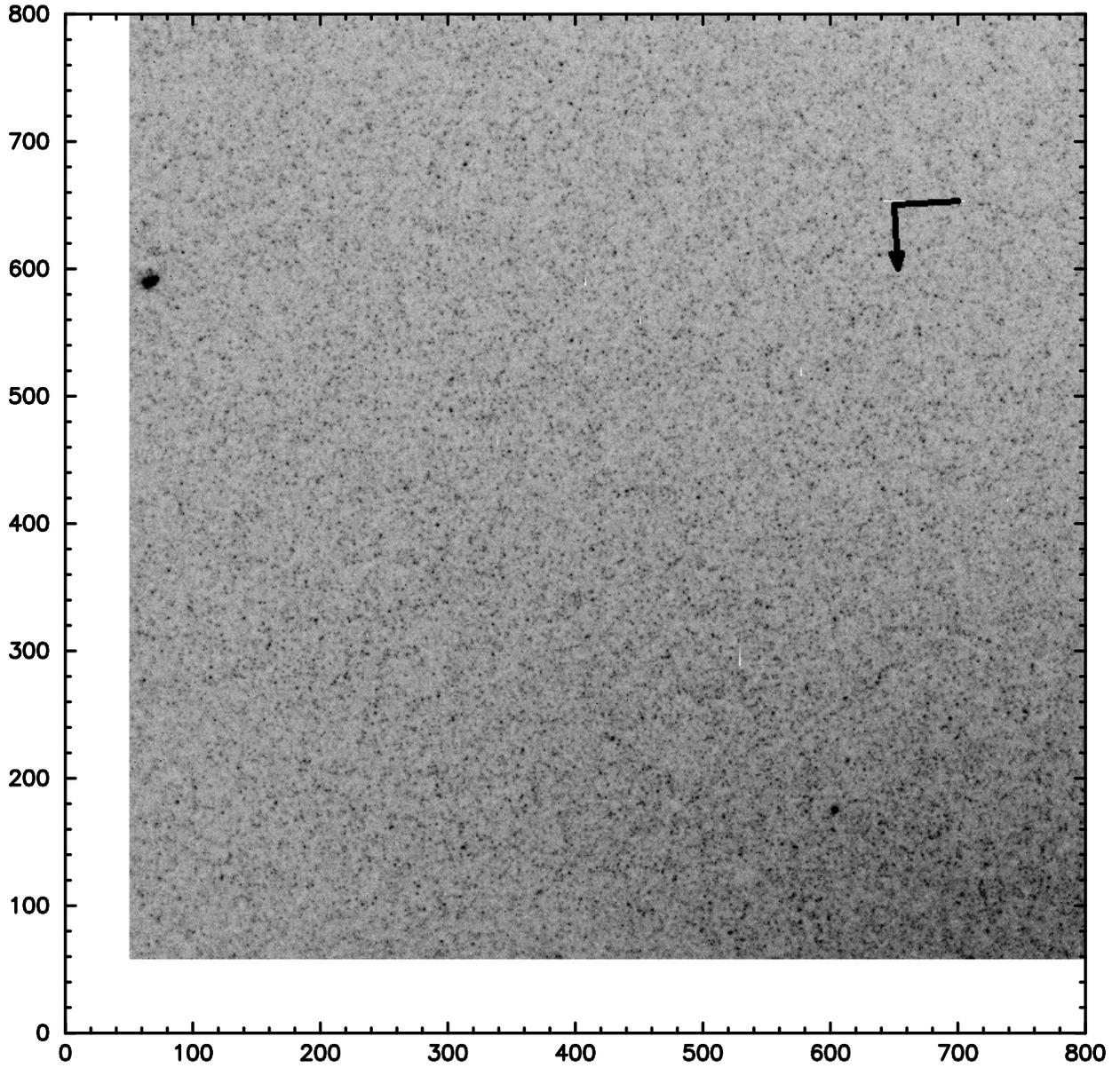}
\caption{{\it HST}/WFPC2 PC image in the $V$- band for the intermediate 
position (field 2) of NGC 5102. The bright resolved stars are visible.
\label{WFPC2-PCs-2}}
\end{figure}

\begin{figure}
\plotone{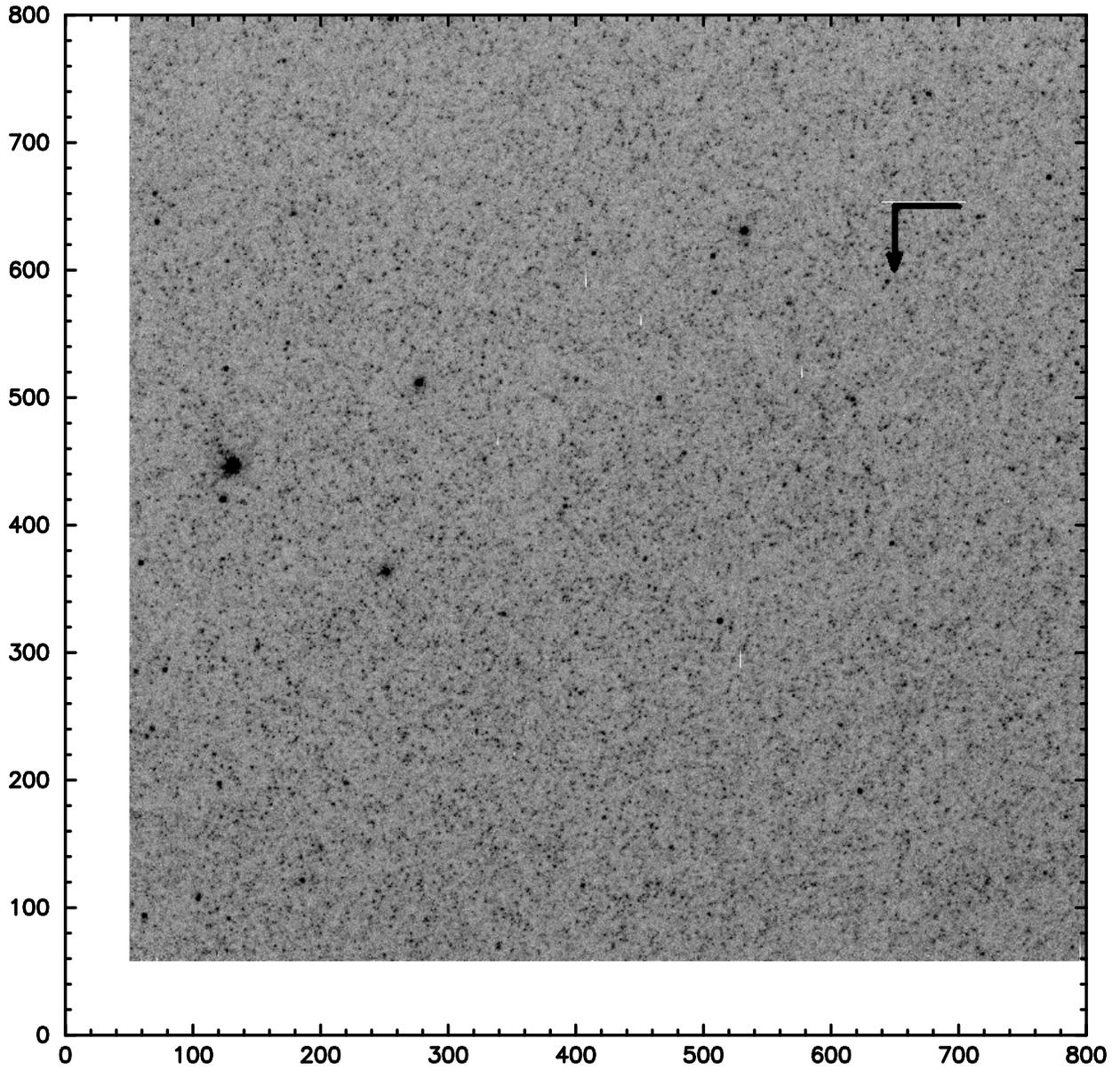}
\caption{{\it HST}/WFPC2 PC image in the $V$- band for the disk field (field 3) 
of NGC 5102.
\label{WFPC2-PCs-3}}
\end{figure}

\begin{figure}
\epsscale{1.0}
\plotone{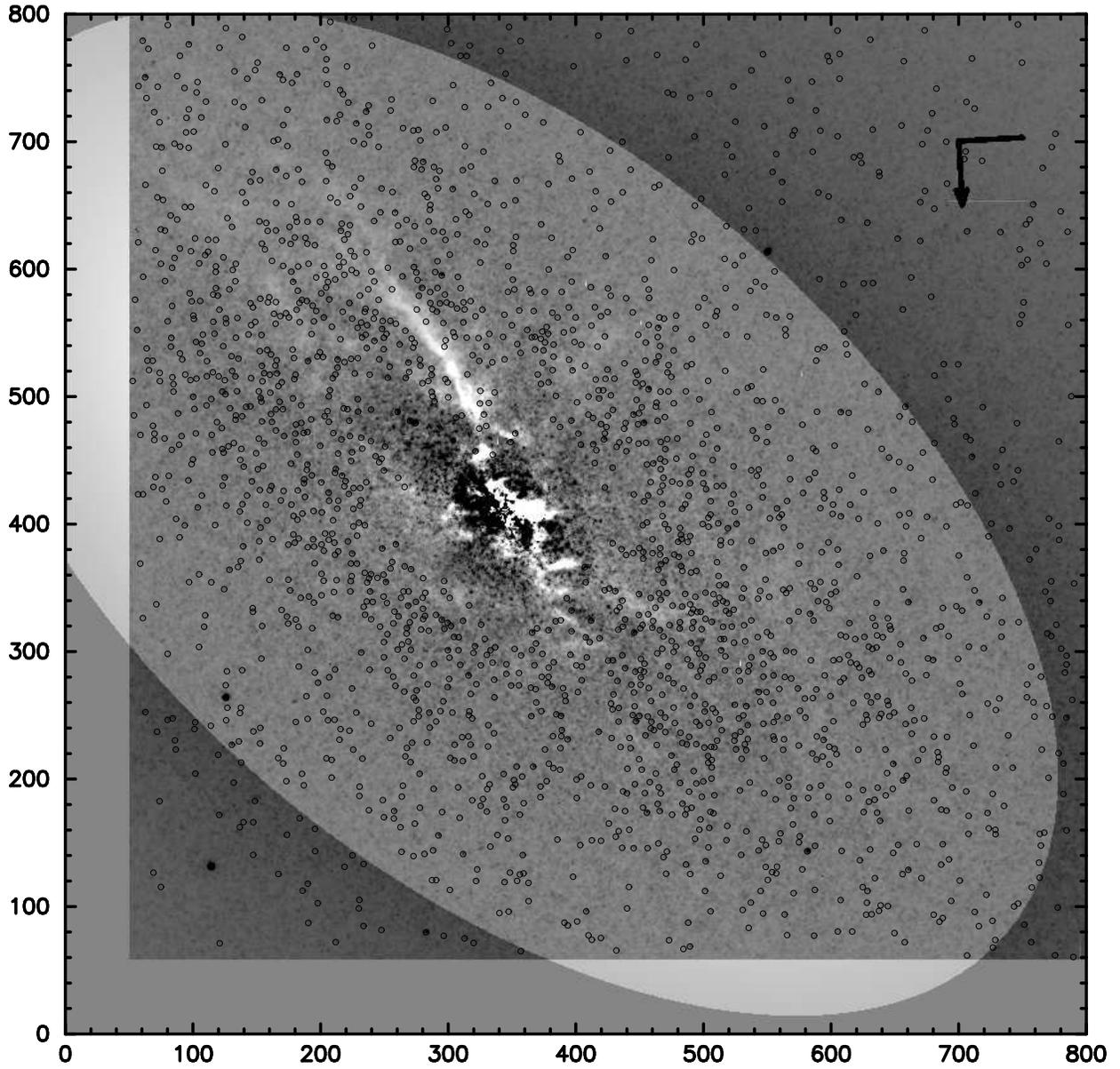}
\caption{{\it HST}/WFPC2 PC negative image of the inner ($12\arcsec$ x $12\arcsec$) 
region of NGC 5102 (field 1) with a smooth model subtracted. The dust lanes and patches, 
as well as the bright resolved stars from the starburst are clearly visible. Overlaid are the 
detected stars. 
\label{BMODEL-stars}}
\end{figure}

\begin{figure}
\epsscale{0.70}
\plotone{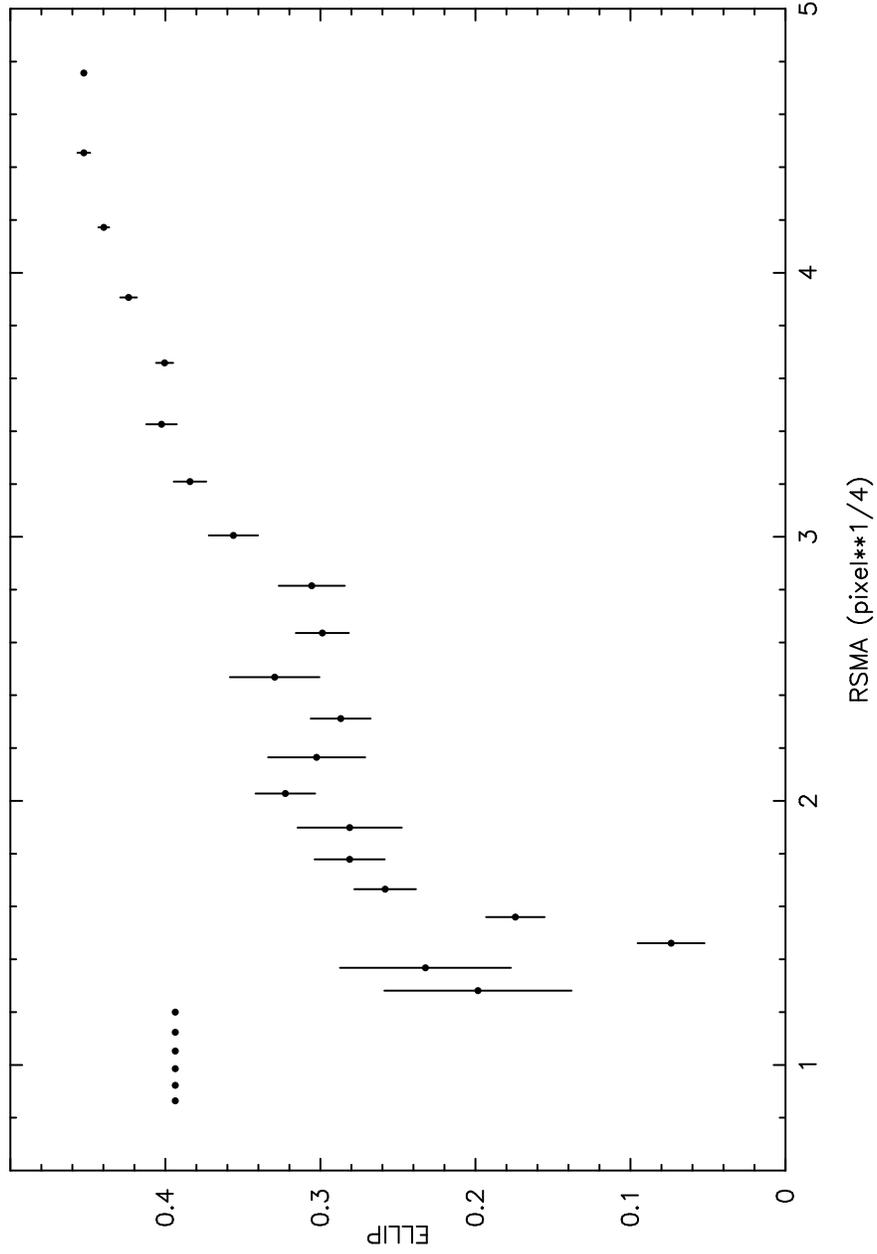}
\caption{{Semimajor axis vs. the ellipticity from the smooth model.}
\label{ELLIP-E}}
\end{figure}

\begin{figure}
\epsscale{0.70}
\plotone{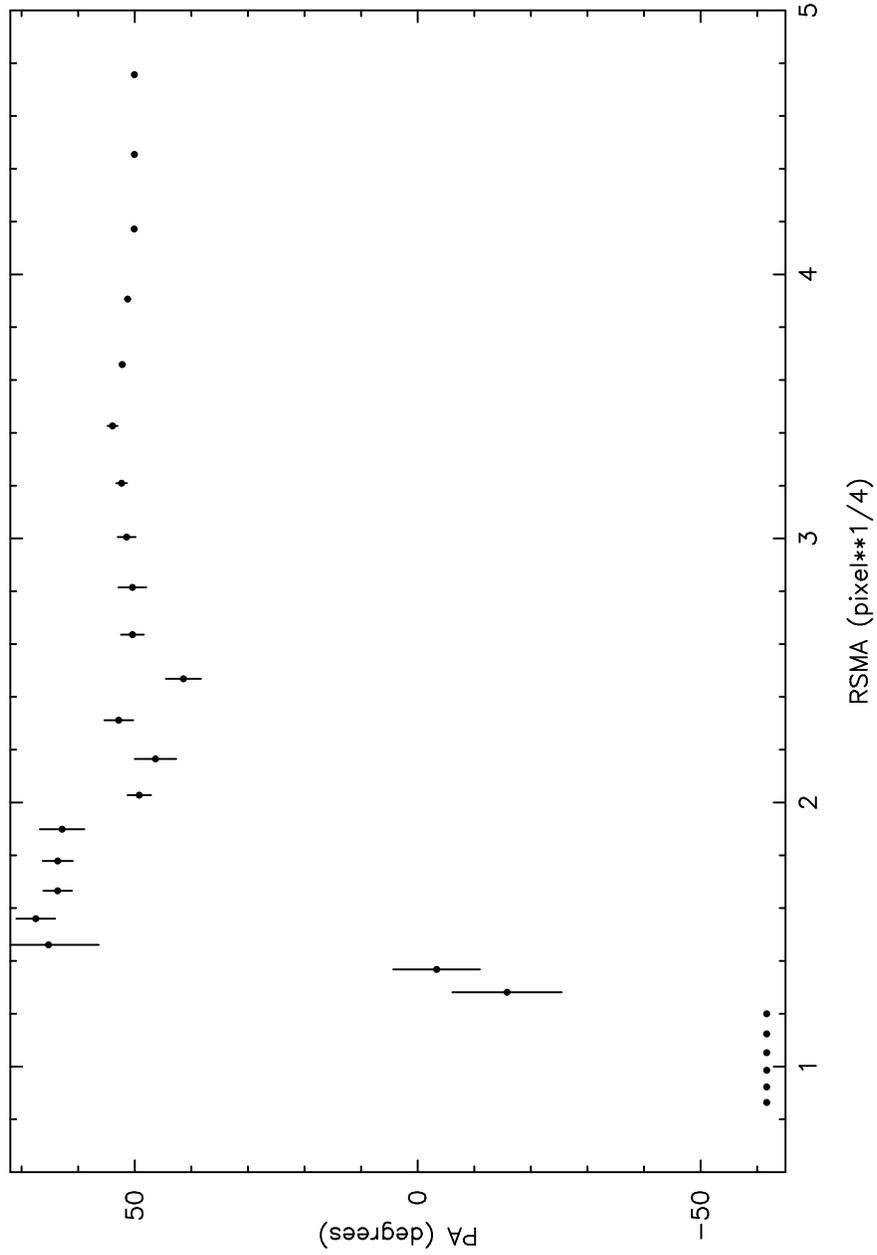}
\caption{{Semimajor axis vs. the position angle from the smooth model. The position angle
of each fitted ellipses (excluding the inner $0\arcsec.5$ radius region) remains fairly constant.}
\label{ELLIP-PA}}
\end{figure}

\begin{figure}
\epsscale{0.80}
\plotone{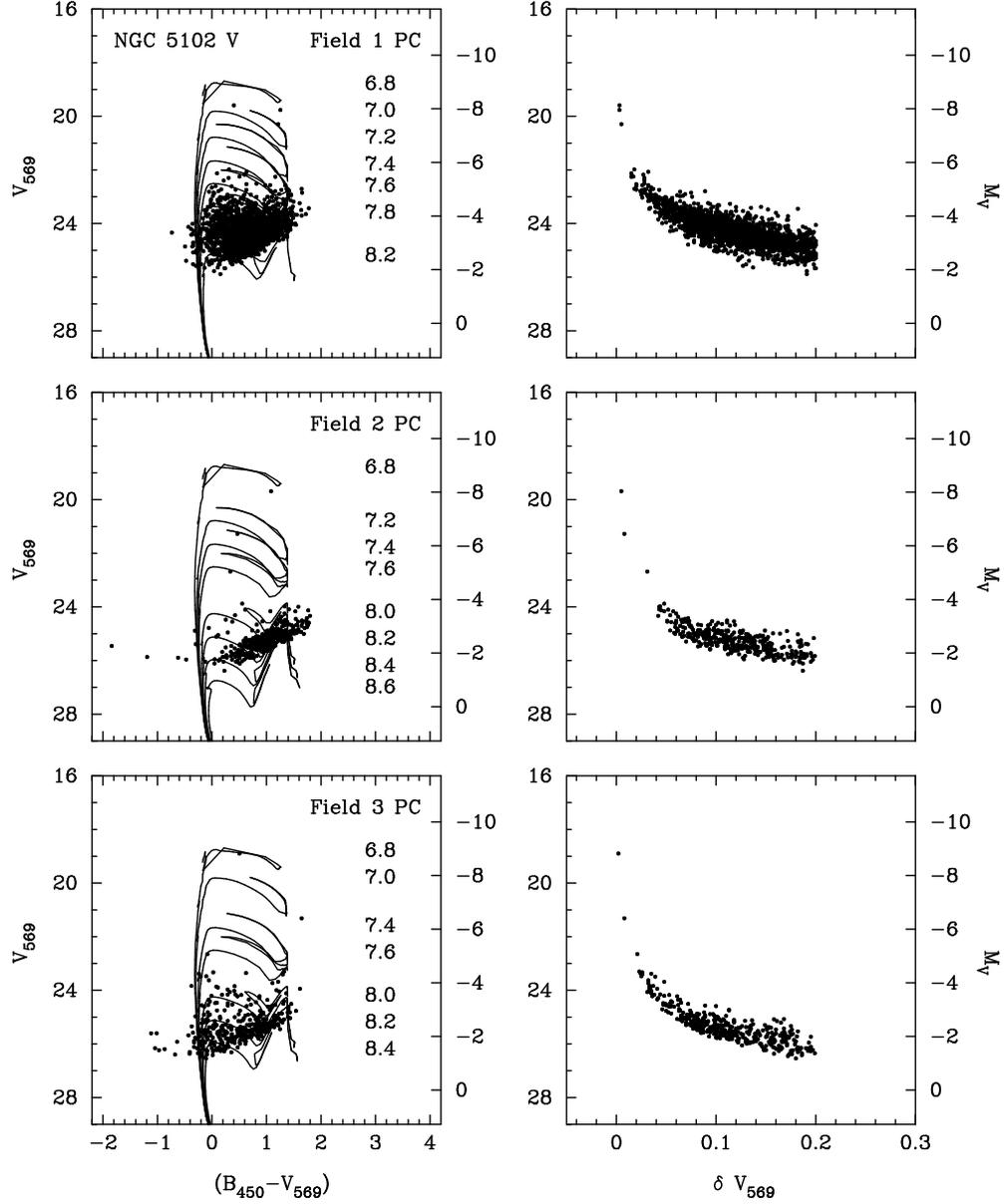}
\caption{Left panel shows colour-magnitude data in the $V$- band for the three fields of 
NGC 5102. Overlaid on the data are $z = 0.02$ Padova isochrones with their log(age).
The Galactic extinction correction has been applied. The right panel shows the associated 
photometric errors.
\label{CMD-PCs-Dered}}
\end{figure}

\begin{figure}
\plotone{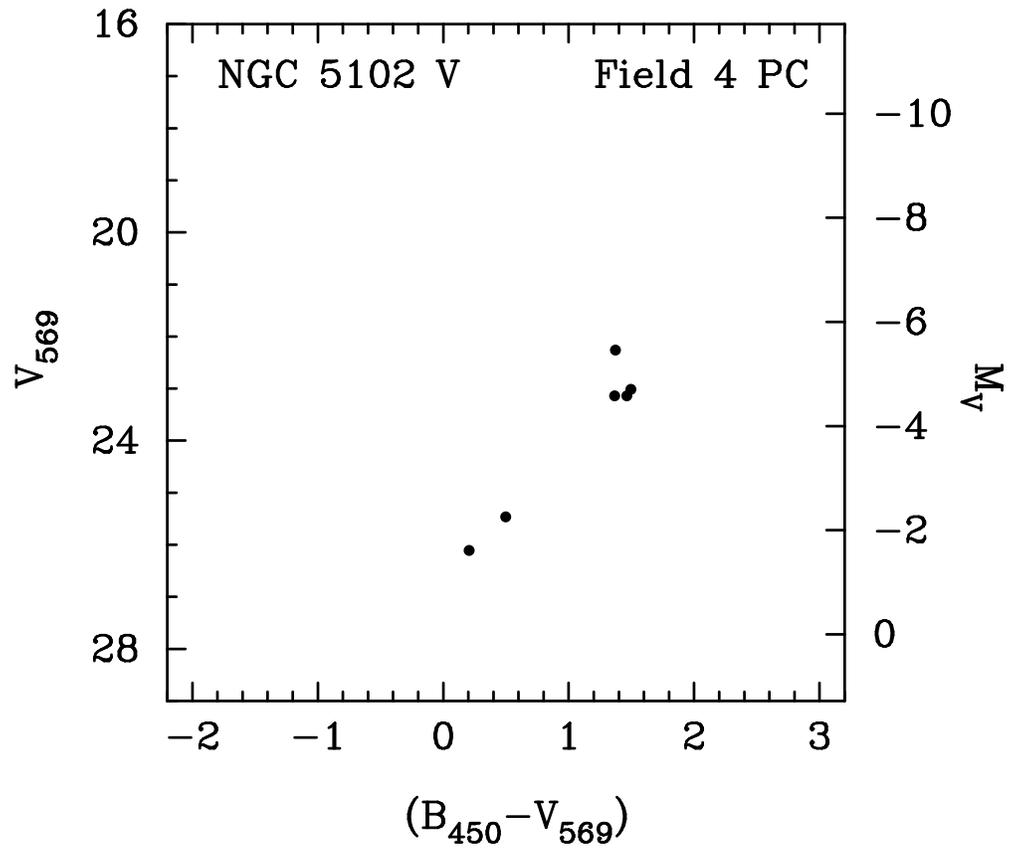}
\caption{Colour-magnitude data in the $V$-band for the background field (field 4).
We note that contamination from background stars is not a problem.
\label{CMD-PCs-bkgd}}
\end{figure}

\begin{figure}
\epsscale{0.80}
\plotone{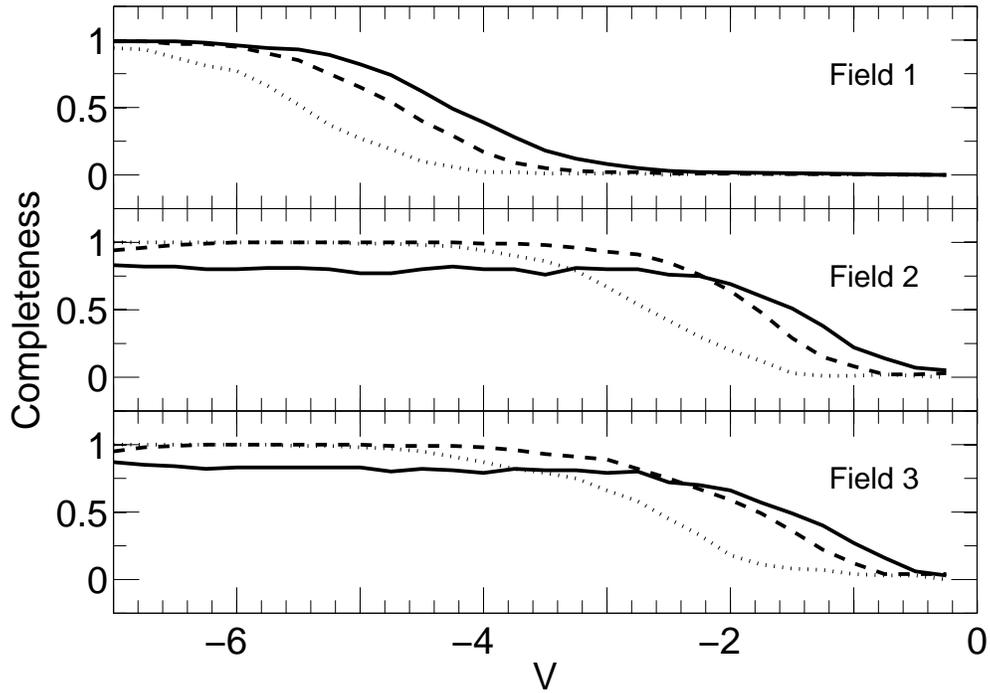}
\caption{Completeness vs. absolute magnitude as a function of colour for the three fields in the 
$V$- band. We see a clear dependence on colour. As expected, the inner field (F1) is more 
crowded than the outer fields (F2 and F3). Solid line is for $(B-V) \leq 0$, dashed line is for 
$0 \leq (B-V) \leq 1$, and dotted line is for  $(B-V) \geq 2$.
\label{CompFunc}}
\end{figure}

\begin{figure}
\epsscale{0.80}
\plotone{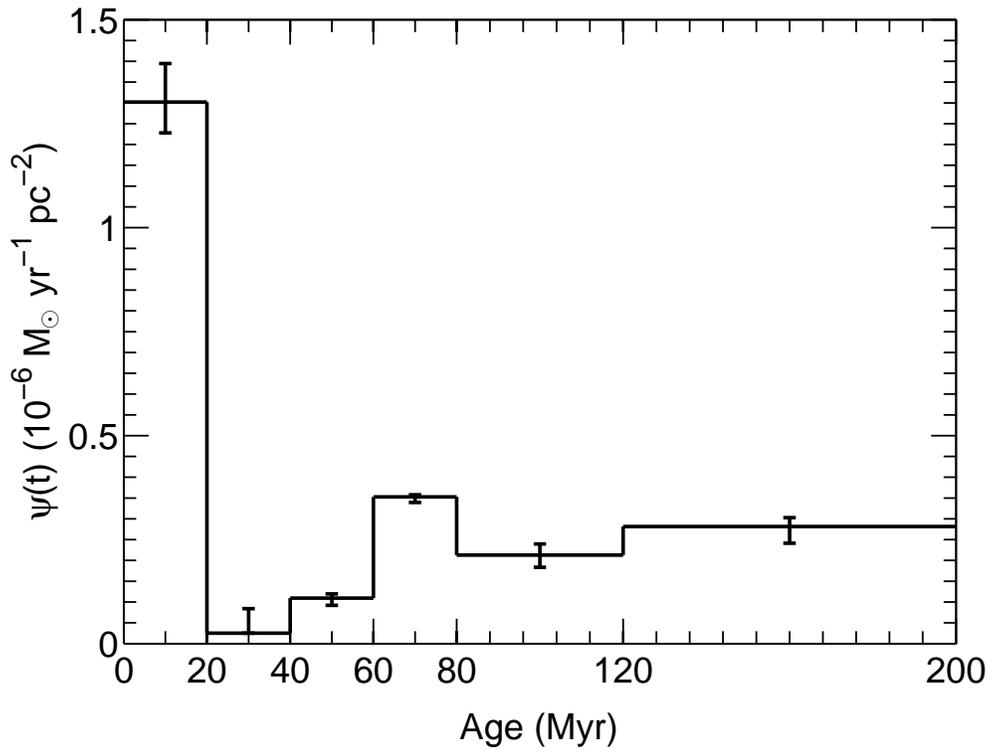}
\caption{SFH of the central field (F1). The bright background of the
central field means that it is not possible to probe the SFH older than
200 Myr. Here, we can see what appears to be a relatively high star formation rate over
the past 20 Myr. The star formation rates shown here (and in Figures 13 and 15) are
per unit area projected on the sky. 
\label{SFH-F1}}
\end{figure}

\begin{figure}
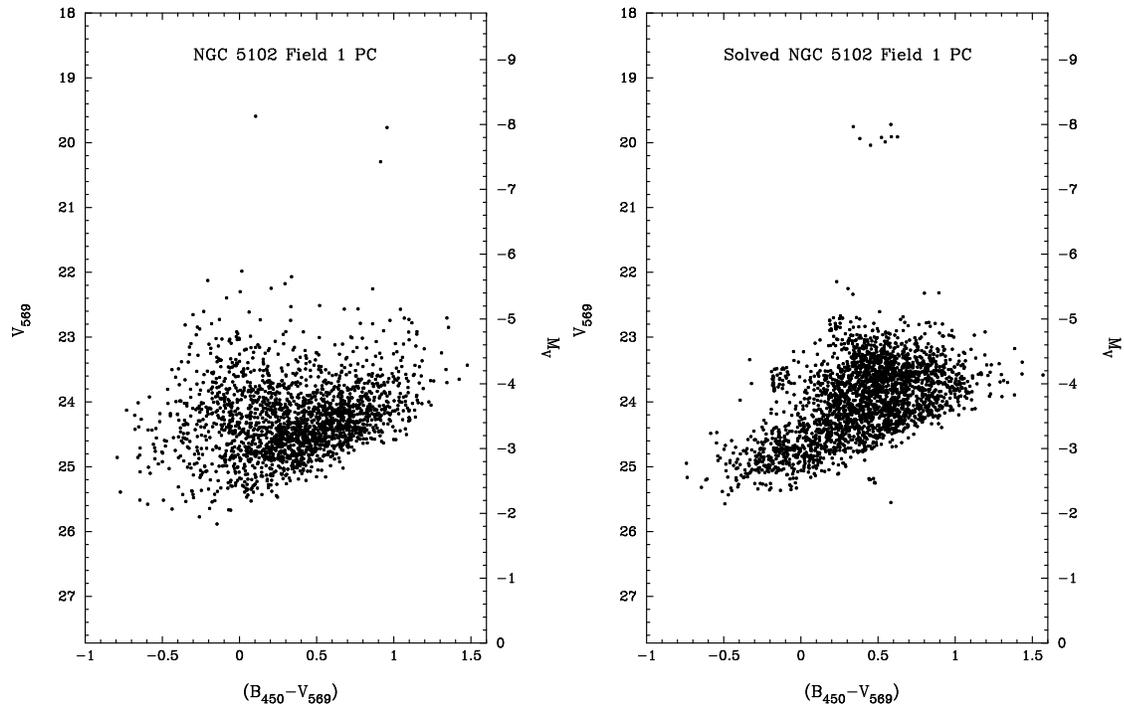

\includegraphics[scale=0.6]{figure12a.ps}
\includegraphics[scale=0.6]{figure12b.ps}
\caption{Observed CMD and solved CMD for the central field (F1).
\label{OBSvsSYN-F1}}
\end{figure}

\begin{figure}
\epsscale{0.80}
\plotone{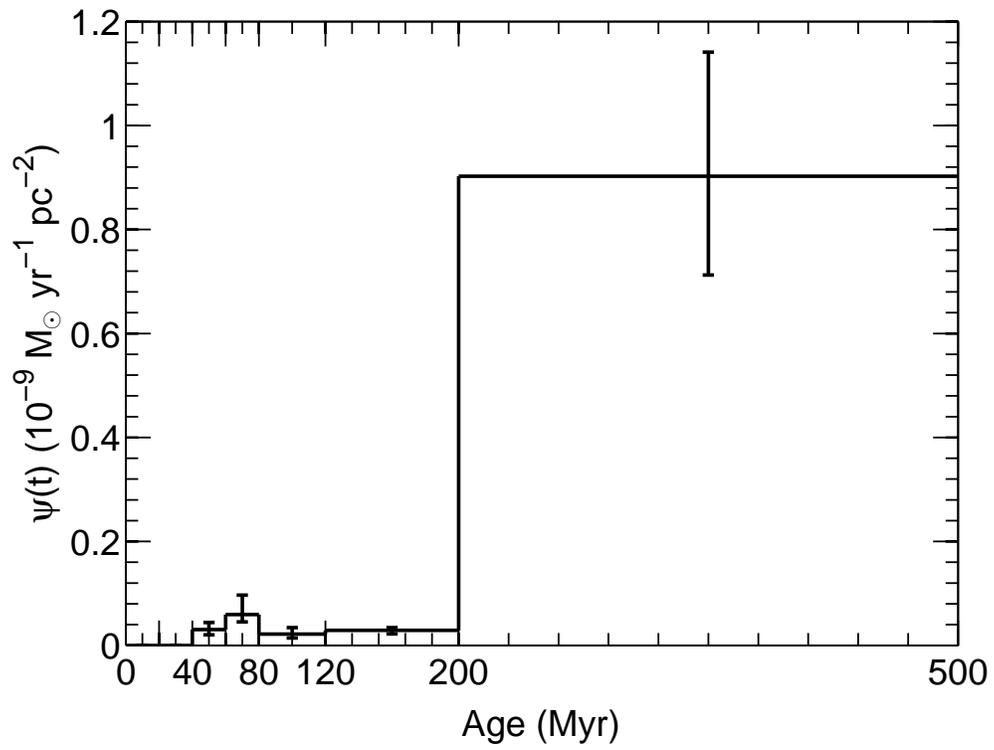}
\caption{SFH of the intermediate field (F2). A decline in the star 
formation is seen from about 0.5 Gyr then, the past 200 Myr appear to have been quiescent. 
\label{SFH-F2}}
\end{figure}

\begin{figure}
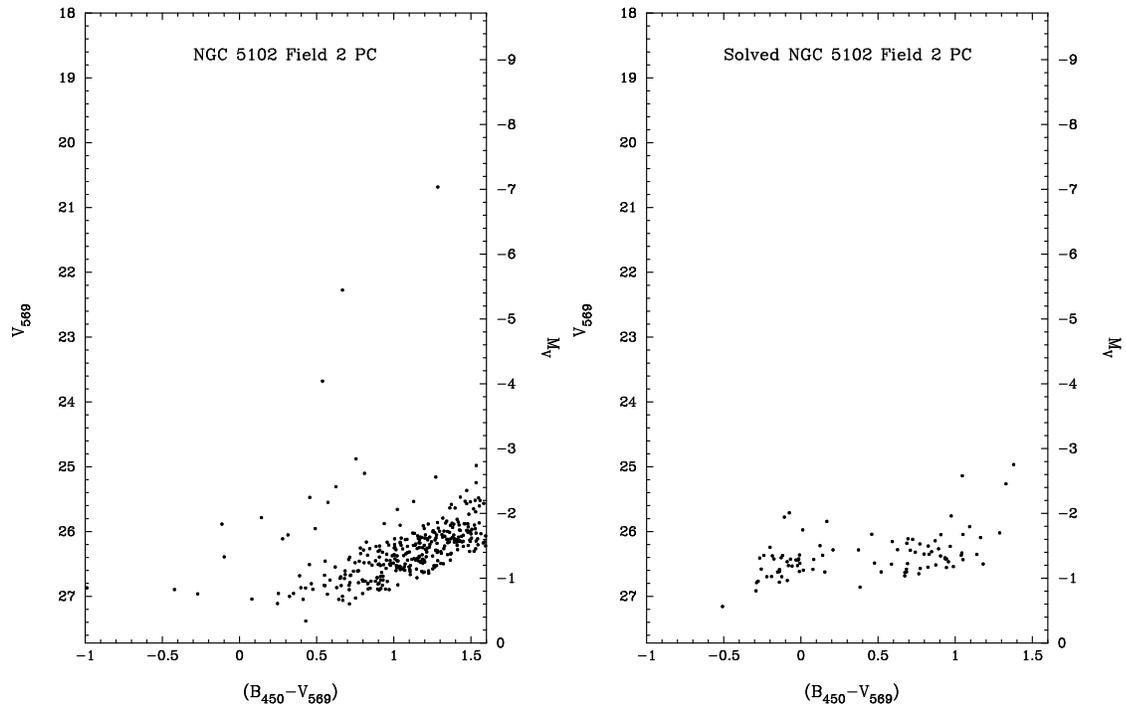

\includegraphics[scale=0.6]{figure14a.ps}
\includegraphics[scale=0.6]{figure14b.ps}
\caption{Observed CMD and solved CMD for the intermediate field (F2).
\label{OBSvsSYN-F2}}
\end{figure}

\begin{figure}
\epsscale{0.80}
\plotone{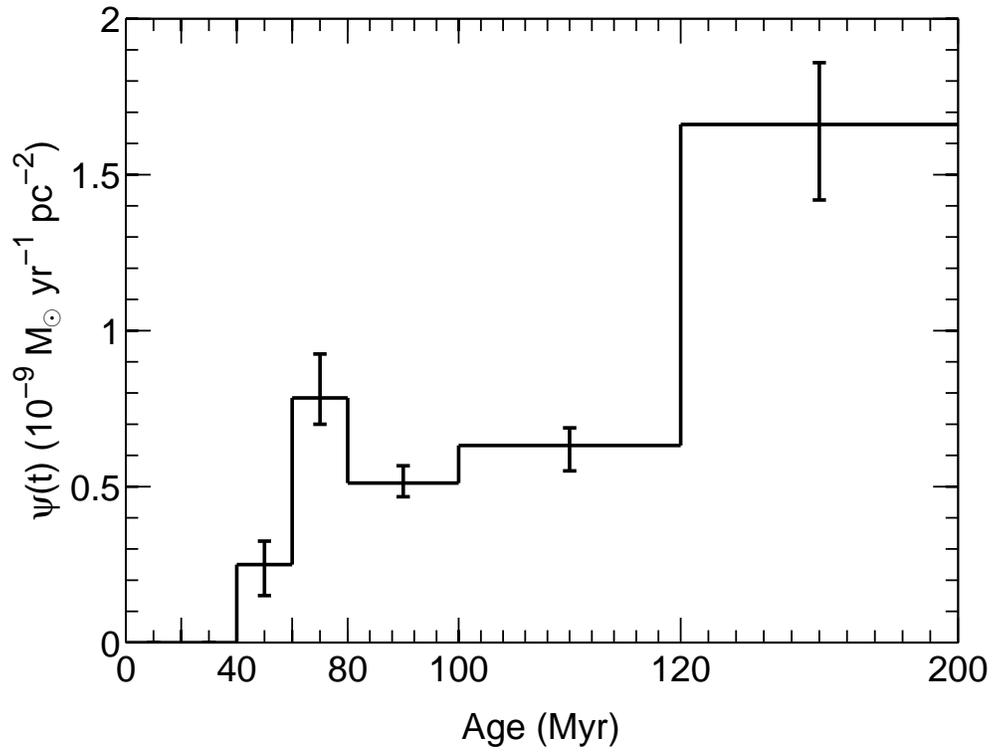}
\caption{SFH of the disk field (F3). The star formation rate declined
smoothly from about 200 Myr ago, and has been quiescent over the past 40 Myr.
\label{SFH-F3}}
\end{figure}

\begin{figure}
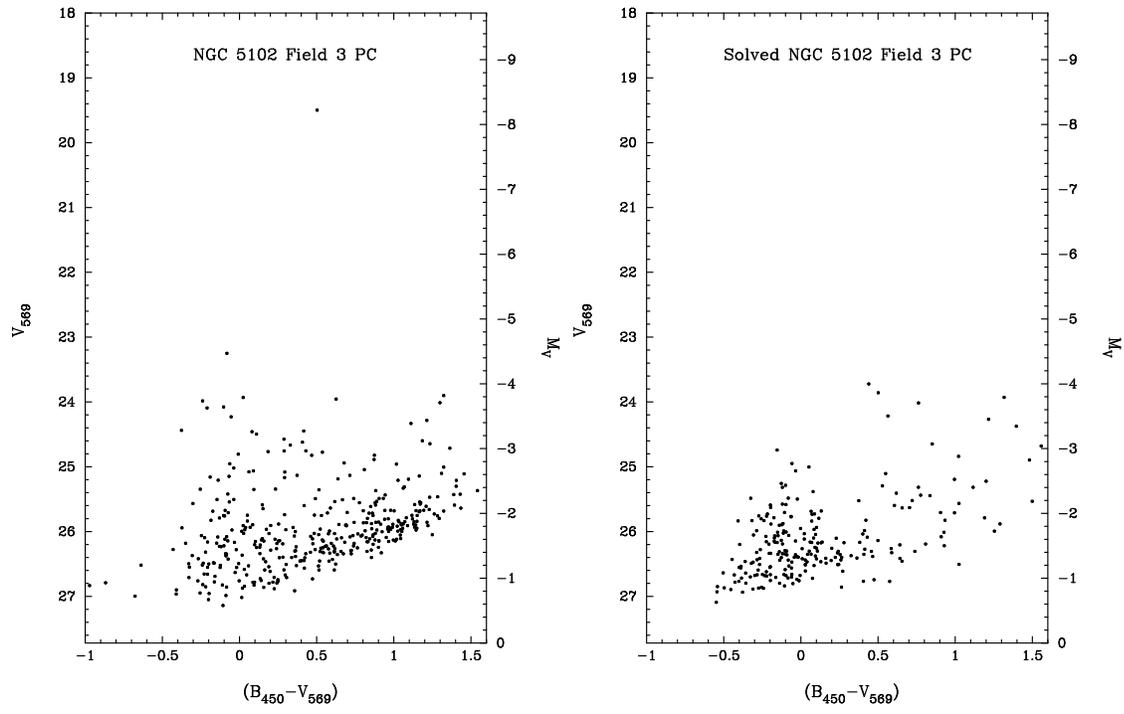

\includegraphics[scale=0.6]{figure16a.ps}
\includegraphics[scale=0.6]{figure16b.ps}
\caption{Observed CMD and solved CMD for the disk field (F3).
\label{OBSvsSYN-F3}}
\end{figure}

For the disk, the mass is about 60\% of the total stellar mass or $3.3 \times 10^9$ \msol. The annuli 
associated with the disk fields F2 and F3 cover about half of the disk area visible in Figure 
\ref {DSS_3POS}. Assuming that the star formation rate in F2 and F3 is typical of the disk, then 
from Table \ref {N5102-sfr} the mass of new stars formed in the disk over the past 200 Myr is about 
$2 \times 10^6$ \msol. This represents less than 0.1\% of the disk mass. We conclude that the star 
formation observed in the disk over the past 200 Myr has contributed negligibly to build up the disk. 
On the other hand, Davidge (2008a) concludes that at least 20\% of the stellar disk mass was formed in 
the past Gyr. If both of these inferences are correct, then it seems clear that the mean star formation 
rate in the period between 200 Myr and 1 Gyr ago was about 80 times higher than the rate which we 
have derived. We are seeing the disk near the end of its recent starburst.

For the bulge, the mass is about 40\% of the total or $2.2 \times 10^9$ \msol. Our F1 covers roughly 
half of the bulge area visible in Figure \ref {DSS_3POS}, so the mass of new stars formed in the bulge 
over the past 200 Myr is about $4 \times 10^7$ \msol. The {\it B-R} colour profile derived by Pritchet (1979)
is approximately constant within the bulge, outside of the nuclear region, so we assume that the star
formation rate per unit area in F1 is representative of the entire bulge area. Then the star formation 
events of the past 200 Myr have contributed about 2\% of the stellar mass of the bulge. 

Although we have no direct information about the star formation rate in the bulge at earlier times, 
this 2\% growth in the last 200 Myr is interesting. It would not take a very large enhancement
of this rate over the previous Gyr to add a substantial amount of stars to the bulge. For example, 
a sixfold increase in the mean star formation rate in the bulge over the period between 200 Myr and 
1 Gyr ago would be enough to double the stellar mass of the bulge in the last Gyr. This sixfold 
increase is much less than the increase in the disk star formation rate over the same period
inferred above. 

Continuing with this argument, we note that the SFH of F1 (Figure \ref{SFH-F1}) shows a 
gradual decay in the star formation rate from 200 Myr ago, a brief enhancement about 70 Myr ago and a
very recent star formation episode in the last 20 Myr. The SFH from 200 Myr to 20 Myr 
ago can be represented by an exponential decay with a timescale of 110 Myr, plus the enhancement at
70 Myr ago. The decaying exponential component contributed about $2.2 \times 10^7$ \msol of stars to 
the bulge over the last 200 Myr, with the remainder coming from the two short-duration bursts. Extrapolating 
this exponential SFH backward in time, we would need to go back only 700 Myr ago for 
this starburst to generate the entire stellar mass of the bulge. If the starburst did indeed start its exponential 
decay 700 Myr ago, the instantaneous star formation rate of the exponential component at that time would 
have been about 20 \msol $\rm yr^{-1}$, compared with the mean rate of about 0.1 \msol $\rm yr^{-1}$ 
over the past 200 Myr.

What is the likely scenario for the evolution of NGC 5102 ?  The colours of the outer disk are similar 
to those of present-day S0 galaxies in the Virgo cluster (see Section 1), indicating that the starburst 
is superimposed on an older disk population. From Davidge's (2008a) argument, the starburst appears 
to have been very substantial, and we can probably exclude the possibility that the residual star
formation seen now is the remains of the normal star formation of a spiral galaxy exhausting its gas 
and making a passive transition to an S0 galaxy.

Before the starburst, NGC 5102 would have been a dwarf disk galaxy with a stellar mass of about 
$4 \times 10^9$ \msol, and an absolute magnitude of about {\it M}$_{\rm B} = -17$. The outer regions 
of NGC 5102 appear to be relatively unaffected by the starburst, and the surface brightness profile 
shown by van Woerden et al. (1993) indicates an outer disk of normal surface brightness (e.g., 
Freeman 1970). NGC 5102 then captured a gas cloud or gas-rich galaxy with an \hi mass of at 
least $2 \times 10^9$ \msol which fueled the starburst and added about 20\% to the existing stellar 
mass, mainly in the bulge. The idea of capture of a gas cloud is not novel in the context of the
Cen A group: similar captures have been invoked to explain the starburst in NGC 5253 and the 
disturbed structure of M83 and Cen A itself (e.g., L\'opez-S\'anchez et al, 2008).  At the peak of
its recent star formation activity, NGC 5102 may have looked rather like the dwarf starburst system 
M82 (e.g., Mayya et al. 2006; Davidge 2008b). Now we see it in a late phase of the star formation event, as an S0 
galaxy with the star formation rate close to zero and the gas almost exhausted: the residual HI mass 
is about $3 \times 10^8$ \msol and the mass of ionised gas is only about $9 \times 10^5$ \msol 
(McMillan et al 1994).

This picture of star formation driven by infalling gas from the observed neutral hydrogen is supported 
by the similarity of dynamical and star formation timescales, all around 30 Myr, for NGC 5102. We list 
some of the relevant timescales below.

(i) McMillan et al (1994) discovered a prominent filament of ionized gas extending roughly from the 
centre of the galaxy in both directions along the major axis of the galaxy which they argue is excited 
by a low velocity shock. The velocity gradient along this feature has an associated timescale of about 
20 Myr, comparable with the age of the most recent star-forming event in our field F1.

(ii) The measured HI rotation curve extends inwards to a radius of 2 kpc. At this radius the rotation 
period is about 160 Myr (van Woerden et al 1993) so the infall time (typically 0.2 $\times$ rotation 
period) is about 30 Myr.

(iii) Although the decay timescale of the star formation rate in F1 is longer (about 110 Myr), the 
SFH shows superimposed events with shorter timescales of about 20 Myr.

(iv) In our disk field (F3), the SFH shown in Figure \ref{SFH-F3} can again be well 
represented by an exponential decay, with a timescale of 58 Myr. As in F1, the SFH 
shows a brief enhancement about 70 Myr ago. The HI infall time at the radius of F3 is about 40 Myr.

There are two important classes of cosmological galaxy formation models that can apply to this interpretation 
of NGC 5102. Both of them may have been active in the case of NGC 5102, and it is worthwhile to discuss 
them here and point out how their relative significance might be determined.

(I) The density and velocity fields of gas in galaxies at redshift $\approx 2$ (where the peak of the star 
formation in the universe occurs) appear clumpy and turbulent. In the remarkable SINFONI observations 
from Genzel and colleagues (Forster-Schreiber et al 2009; Shapiro et al. 2009; Burkert et al. 2009) one 
sees a small number of massive clumps. Now, the relaxation time for a clumpy disk with {\it N} clumps is 
$T_R \sim NT_d$ where $T_d$ is the dynamical time. In this picture then, once a clump is released from 
the HI disk and sent toward the central region, the timescale would be $T_d$ and the time in between such 
events would be $T_R \sim 3 -10~ T_d$ (assuming there are 3 -10 clumps). Thus, a secular bulge building 
process can occur in such disks. Recent interesting theoretical models of Birnboim et al (2007), Dekel et al 
(2009a \& 2009b), and Ceverino et al (2009) may also apply to cold (HI) clumpy turbulent HI distributions 
and flows as discussed here.

(II) The classic starburst feedback ideas that apply to downsizing and quenching of the galaxy formation 
process in many available feedback models involve halting the formation of galaxies by feedback processes. 
In the case of NGC 5102, one would imagine the following cycle. The HI distribution builds up in the centre 
and then a threshold is reached where at least part of the central HI mass becomes dynamically unstable 
and a starburst results on the dynamical timescale. Then a starburst wind (or in some cases an AGN 
generated feedback process) is triggered, the remaining gas is then blown away and the star formation process 
is halted. The central region is then replenished on a timescale which is something like the turbulent-velocity crossing time of the central region $-$ in the case of NGC 5102, one has a turbulent velocity of say 
$V_t \sim$ 10 km s$^{-1}$ and a scale of $R_B \sim 1$ kpc, resulting in a refilling time - during which the 
star formation is effectively quenched - of  $T_q  \sim 100$ Myr. We do not see the feedback process of the 
starburst driven wind in NGC 5102, but we do see the large scale ionised gas shell that it may have formed.

How to estimate the relative importance of these two scenarios in the case of NGC 5102 ?
(1) The HI map needs to be repeated with significantly higher spatial resolution. The question of HI 
clumpiness is important to resolve. Similarly, molecular (CO) observations are also essential.
(2) The details of the possible feedback from the shell structure and other associated gas are now most 
interesting and new deep X-ray and spectroscopic studies would be the best way to proceed.
(3) Detailed kinematic data on the bulge of NGC 5102 might help determine if there were randomly 
distributed blobs falling in to make the bulge, or a more quiescent picture of central disk refilling.

Sorting out the relationship between the higher redshift, $z \sim 2$, observations and this excellent ÒlocalÓ 
laboratory for bulge formation provides a rare opportunity

It would be interesting to know whether the bulge of NGC 5102 was primarily built during the recent 
starburst. To the level at which we can work here, this cannot be excluded. As a test of this possibility, 
it would be useful to search for older stars in the bulge: this is probably best done with adaptive optics 
or {\it HST} imaging in the near-IR, to detect the tip of the old ($> 1$ Gyr) red giant branch against the bright 
background of the bulge. A negative outcome of such a search for old bulge stars in NGC 5102 would 
suggest that its bulge was indeed primarily built up in this one recent and brief star-forming event. 

Such rapid formation of a bulge at a relatively recent epoch would be a different mode of bulge formation 
from the more usual scenarios of forming classical and pseudo bulges in disk galaxies (e.g. Kormendy 
\& Kennicutt 2004). NGC 5102 is a low-luminosity S0 system. Even if its bulge turns out to have 
formed in a single recent event, one cannot generalise its SFH to more luminous S0 
galaxies in general. However we are tempted to ask whether this kind of rapid bulge formation, driven 
by infall and starbursts at later and maybe multiple epochs, could be more common in S0 galaxies with 
their relatively quiescent disks. If so, it may be at least part of the reason why S0 galaxies have 
predominantly larger bulges than spirals (e.g.,  Dressler 1980).

\acknowledgments

We are grateful to Leo Girardi for providing the unpublished set of isochrones for {\it HST}/WFPC2 
F569W. SFB thank J. Gallagher and C. Pritchet for fruitful discussions. We thank the
referee for helpful comments which improved the presentation of the paper. This study was partly 
funded by NASA grant HST-GO-07455.01-96A. This research has made use of the NASA/IPAC
Extragalactic Database (NED) which is operated by the Jet Propulsion Laboratory, California Institute
of Technology, under contract with the National Aeronautics and Space Administration. This research
has made use of NASA's Astrophysics Data System Bibliographic Services, and of photographic data
obtained using the UK Schmidt Telescope. The UK Schmidt Telescope was operated by the Royal 
Observatory Edinburgh, with funding from the UK Science and Engineering Research Council, until 1988
June, and thereafter by the Anglo-Australian Observatory. The Digitized Sky Survey was produced at the 
Space Telescope Science Institute under US Government grant NAG W-2166. To produce some of the 
figures, we used the KARMA package (Gooch 1997).

\clearpage
\newpage


\begin{deluxetable}{lcc}
\tablecolumns{3}
\tablewidth{0pt}
\tablecaption{Basic Parameters for NGC 5102 \label{N5102-param}}
\tablehead{
\colhead{Parameter}  & \colhead{Value} & \colhead{Notes}}
\startdata
R.A. (J2000.0) & $13~21~57.6$ & 1 \\
decl. (J2000.0)  & $-36~37~49$ & 1 \\
Morphological type & SA0 & 1 \\
Adopted distance (Mpc) & 3.5 & \\
Distance modulus $(m-M)_o$ & 27.72 & \\
Helio radial velocity ($\rm km~ s^{-1}$) & 468 & 1,2 \\
Major diameter (arcmin) & 8.7 & 1 \\
Minor diameter (arcmin) & 2.8 & 1 \\ 
Inclination (deg) & 66 &  \\
P.A. major axis (deg) & 48 & 3 \\
$\rm M_{HI}$ ~(\msol) & $2 \times 10^8$ & 4 \\
\mhil ~(\mlsol) & 0.12 & 4 \\
$\rm [Fe/H]$ & 0.0 & 5 \\
Reddening $E(B-V)$ & 0.055 & 1 \\
Total $B$ magnitude $\rm B_T$ & 10.35 & 3 \\
$\rm (B-V)_T$ & 0.72 & 3 \\
\enddata
\tablerefs{(1) NED; (2) Koribalski et al. 2004; (3) RC3; (4) van Woerden et al. 1993 with 
the $\rm M_{HI}$ adjusted to our adopted distance; (5) refers to the young population}
\end{deluxetable}
\clearpage
\newpage

\begin{deluxetable}{ccccc}
\tablecolumns{5}
\tablewidth{0pt}
\tablecaption{{\it HST}/WFPC2 Data Sets for NGC 5102 : {\it HST} Program 5400
\label{N5102-obs}}
\tablehead{
\colhead{Data Set}  &
\colhead{Filter}        & 
\colhead{Field}        &
\colhead{Date}        & 
\colhead{Exposure (s)} }
\startdata
u2bt0101t & F450W & 1 & 1994 Sep 2 & 400  \\
u2bt0102t & F450W & 1 & 1994 Sep 2 & 400  \\
u2bt0103t & F450W & 1 & 1994 Sep 2 & 400  \\
u2bt0104t & F450W & 1 & 1994 Sep 2 & 400  \\
u2bt0105t & F450W & 1 & 1994 Sep 2 & 400  \\
u2bt0106t & F450W & 1 & 1994 Sep 2 & 400  \\
u2bt0107t & F450W & 1 & 1994 Sep 2 & 400  \\
u2bt0108t & F450W & 1 & 1994 Sep 2 & 400  \\
u2bt0109t & F569W & 1 & 1994 Sep 2 & 500  \\
u2bt010at & F569W & 1 & 1994 Sep 2 & 500  \\
u2bt010bt & F569W & 1 & 1994 Sep 2 & 500  \\
u2bt010ct & F569W & 1 & 1994 Sep 2 & 500  \\
u2bt010dt & F569W & 1 & 1994 Sep 2 & 500  \\
u2bt010et & F569W & 1 & 1994 Sep 2 & 500  \\
u2bt010ft & F569W & 1 & 1994 Sep 2 & 500  \\
u2bt010gt & F569W & 1 & 1994 Sep 2 & 500  \\
u2bt010ht & F450W & 2 & 1994 Sep 2 & 1100 \\
u2bt010it & F450W & 2 & 1994 Sep 2 & 1100 \\
u2bt010jt & F450W & 2 & 1994 Sep 2 & 1100 \\
u2bt010kt & F569W & 2 & 1994 Sep 2 & 1100 \\
u2bt010lt & F569W & 2 & 1994 Sep 2 & 1100 \\
u2bt010mt & F569W & 2 & 1994 Sep 2 & 1100 \\
u2bt0201t & F450W & 3 & 1994 Sep 8 & 1100 \\
u2bt0202t & F450W & 3 & 1994 Sep 8 & 1100 \\
u2bt0203t & F450W & 3 & 1994 Sep 8 & 1100 \\
u2bt0204t & F569W & 3 & 1994 Sep 8 & 1100 \\
u2bt0205t & F569W & 3 & 1994 Sep 8 & 1100 \\
u2bt0206t & F569W & 3 & 1994 Sep 8 & 1100 \\
u2bt0301t & F450W & 4 & 1995 Mar 20 & 1100 \\
u2bt0302t & F450W & 4 & 1995 Mar 20 & 1100 \\
u2bt0303t & F450W & 4 & 1995 Mar 20 & 1100 \\
u2bt0304t & F569W & 4 & 1995 Mar 20 & 1100 \\
u2bt0305t & F569W & 4 & 1995 Mar 20 & 1100 \\
u2bt0306t & F569W & 4 & 1995 Mar 20 & 1100 \\
\enddata
\end{deluxetable}
\clearpage
\newpage

\begin{deluxetable}{cccccc}
\tablecolumns{6}
\tablewidth{0pt}
\tablecaption{{\it HST}/WFPC2 PC Positions \label{N5102-pos}}
\tablehead{
\colhead{Field}                           & 
\colhead{R.A., Decl. (J2000.0)} &
\colhead{Position}                      &
\colhead{P.A.}                            & 
\colhead{Total Exp. (s)} }
\startdata
1 & $13~21~57.9~ -36~37~49.0$ & centre & $176^{o}.76$ & F450W = 3200 \\
 & & & & F569W = 4000 \\
2 & $13~21~54.2~ -36~38~39.1$ & $66\arcsec$ SW& $176^{o}.76$ & F450W = 3300 \\
 & & & & F569W = 3300 \\
3 & $13~21~49.9~ -36~39~20.1$ & $132\arcsec$ SW & $180^{o}.56$ & F450W = 3300 \\
 & & & & F569W = 3300 \\
4 & $13~22~23.7~ -36~37~18.4$ & Backgrd & & F450W = 3300 \\
 & & & & F569W = 3300 \\
\enddata
\end{deluxetable}
\clearpage
\newpage

\begin{deluxetable}{cccl}
\tablecolumns{4}
\tablewidth{0pt}
\tablecaption{Derived Star Formation Densities and Masses \label{N5102-sfr}}
\tablehead{
\colhead{Field}  &
\colhead{Mean SFD}  &
\colhead{Stellar Mass} &
\colhead{Time Interval} \\
\colhead{}  &
\colhead{(\msol pc$^{-2}$)} &
\colhead{(\msol)}  &
\colhead{(Myr)} }
\startdata
F1 & 30 & 2.1$ \times 10^7$  & 0 - 200  \\
F2 & 0.003 & 1.3$ \times 10^4$ & 0 - 200  \\
F3 & 0.11 & 1.0$ \times 10^6$ & 0 - 200  \\
\enddata
\tablecomments{Note. See the text for explanation}
\end{deluxetable}

\clearpage
\newpage


\clearpage
\newpage

\clearpage
\newpage

\clearpage
\newpage

\clearpage
\newpage

\clearpage
\newpage

\clearpage
\newpage

\clearpage
\newpage

\clearpage
\newpage

\clearpage
\newpage

\clearpage
\newpage

\clearpage
\newpage

\clearpage
\newpage

\clearpage
\newpage

\clearpage
\newpage

\clearpage
\newpage

\begin{references}
\reference{} Aparicio, A., \& Gallart, C. 2004, \aj, 128, 1465
\reference{} Aparicio, A., \& Hidalgo, S.L. 2009, AJ, 138, 558
\reference{} Bell, E.F. \& de Jong, R.S. 2001, ApJ, 550, 212
\reference{} Bendo, G. J., \& Joseph, R. D. 2004, \aj, 127, 3338
\reference{} Biretta, J., et al. 1996, WFPC2 Instrument Handbook (version 4.0;
Baltimore:STScI)
\reference{} Birnboim, Y., Dekel, A., \& Neistein, E. 2007, MNRAS 380, 339
\reference{} B\"oker, T., Sarzi, M., McLaughlin, D. E., van der Marel, R. P., Rix, H.-W., Ho, L. C., Shields, J. C.
2004, \aj, 127, 105
\reference{} Burkert, A. et al.2009 arXiv 0907.4777
\reference{} Ceverino, D., Dekel, A. \& Bournaud, F. 2009, ArXiv 0907.3271
\reference{} Chandar, R., Leitherer, C., \&  Tremonti, C. A. 2004, \apj, 604, 153
\reference{} Danks, A. C., Lausten, S. \& van Woerden, H. 1979, \aap, 73, 247
\reference{} Davidge, T.J. 2008a, \aj, 135, 1636
\reference{} Davidge, T.J. 2008b, \aj, 136, 2502
\reference{} de Blok, W.J.G., Zwaan, M.A., Dijkstra, M., Briggs, F.H. \& Freeman, K.C. 2002
A\&A, 382, 43
\reference{} Dekel, A., Sari, R., \& Ceverino, D. 2009b, 2009b, ApJ, 703, 785
\reference{} Dekel, A. et al  2009a, Nature 457, 451
\reference{} Dolphin, A.E. 2000a, \pasp, 112, 1383
\reference{} Dolphin, A.E. 2000b, \pasp, 112, 1397
\reference{} Dressler, A. 1980, \apj, 236, 351
\reference{} Forster-Schreiber, N.M. et al 2009, ApJ, 706, 1364
\reference{} Freeman, K. 1970, \apj, 160, 811
\reference{} Gallagher, J., Faber, S., \& Balick, B. 1975, \apj, 202, 7
\reference{} Girardi, L., Bertelli, G., Bressan, A., Chiosi, C., Groenewegen, M.A.T.,
Marigo, P., Salasnich, B., \& Weiss, A. 2002, \aap, 391, 195
\reference{} Girardi, L., Bressan, A., Bertelli, G., \& Chiosi, C. 2000, A\&AS, 141, 371 
\reference{} Gooch, R.E. 1997, PASA, 14, 106
\reference{} Irwin, J.A., Bregman, J.N., Athey, A. E.  2004, \apj, 601, 143
\reference{} Karachentsev, I.D, et al. 2002, \aap, 385, 21
\reference{} Koribalski et al. 2004, AJ, 128, 16
\reference{} Kormendy, J. \& Kennicutt, R.C. 2004, ARAA, 42, 603
\reference{} Kroupa, P., Tout, C.A., \& Gilmore, G. 1993, MNRAS, 262, 545
\reference{} L\'opez-S\'anchez, A., Koribalski, B.,Esteban, C., Garcia-Rojas, J. 
2008, Galaxies in the Local Volume, Astrophysics and Space Science 
Proceedings, Volume . ISBN 978-1-4020-6932-1. Springer Netherlands, 
2008, p. 53
\reference{} Mayya, Y.D., Bressan, A., Carrasco, L., \& Hernadez-Martinez, L. 2006, ApJ, 649, 172
\reference{} McMillan, R., Ciardullo, R., \& Jacoby, G. H. 1994, \aj, 108, 1610
\reference{} Origlia, L. \& Leitherer, C. 2000, AJ, 119, 2018
\reference{} Persson, S.E., Frogel, J.A. \& Aaronson, M. 1979, ApJS, 39, 61
\reference{} Pritchet, C. 1979, \apj, 231, 354
\reference{} Rocca-Volmerange, B., \& Guiderdoni, G. 1987, \aap, 175, 15
\reference{} Sandage, A. \& Visvanathan, N. 1978, \apj, 273, 707
\reference{} Schlegel, D.J., Finkbeiner, D.P., \& Davis, M. 1998, \apj, 500, 525
\reference{} Shapiro, K. et al. 2009, 7 pages, to appear in the proceedings of "Galaxy Evolution: 
Emerging Insights and Future Challenges," Austin, TX, 11-14 Nov 2008
\reference{} Tremonti, C.S., et al. 2004, ApJ, 613, 898
\reference{} van den Bergh, S. 1976, \aj, 81, 795
\reference{} van Woerden, H., van Driel, W., Braun, R., \& Rots, A. H. 1993, \aap, 269, 15
\reference{} Williams, T. B., \& Schwarzschild, M. 1979, \apj, 227, 56
\reference{} Xilouris, E. M., Madden, S. C., Galliano, F., Vigroux, L., \& Sauvage, M. 2004,
\aap, 416, 41
\end{references}
\end{document}